\documentclass[aps,prb,twocolumn,superscriptaddress,showpacs]{revtex4}
\usepackage{graphicx}

\begin{document}
\title{Colossal magnon-phonon coupling in multiferroic Eu$_{0.75}$Y$_{0.25}$MnO$_3$.}

\author{R. Vald\'{e}s Aguilar}
\affiliation{Materials Research Science and Engineering Center,
University of Maryland, College Park, Maryland 20742}

\author{A. B. Sushkov}
\affiliation{Materials Research Science and Engineering Center,
University of Maryland, College Park, Maryland 20742}

\author{C. L. Zhang}
\affiliation{Rutgers Center for Emergent Materials and Department of
Physics \& Astronomy, Rutgers University, Piscataway, New Jersey
08854}

\author{Y.J. Choi}
\affiliation{Rutgers Center for Emergent Materials and Department of
Physics \& Astronomy, Rutgers University, Piscataway, New Jersey
08854}

\author{S.-W. Cheong}
\affiliation{Rutgers Center for Emergent Materials and Department of
Physics \& Astronomy, Rutgers University, Piscataway, New Jersey
08854}

\author{H. D. Drew}
\affiliation{Materials Research Science and Engineering Center,
University of Maryland, College Park, Maryland 20742}



\begin{abstract}
We report the spectra of magnetically induced electric dipole
absorption in Eu$_{0.75}$Y$_{0.25}$MnO$_3$ from temperature
dependent far infrared spectroscopy (10-250 cm$^{-1}$). These
spectra, which occur only in the $e||a$ polarization, consist
of two relatively narrow electromagnon features that onset at
$T_{FE}=30$~K and a broad absorption band that persists to
temperatures well above $T_N=47$~K. The observed excitations
account for the step up of the static dielectric constant in
the ferroelectric phase. The electromagnon at 80 cm$^{-1}$ is
observed to be strongly coupled to the nearby lowest optical
phonon which transfers more than $1/2$ of its spectral weight
to the magnon. We attribute the origin of the broad background
absorption to the two magnon emission decay process of the
phonon.
\end{abstract}

\pacs{
63.20.Ls, 
75.50.Ee, 
75.30.Et, 
76.50.+g 
}

\maketitle

In multiferroic materials the simultaneous magnetic and
ferroelectric order can produce cross coupling between electric
and magnetic signals \cite{Cheong-Mostovoy}. This prospect has
important implications for electronic memory and logic
applications. On the other hand multiferroics exhibit interesting
new fundamental features.  The absence of both time and space
inversion symmetry in relatively low symmetry crystal structures
can produce a rich array of novel magnetoelectric phenomena.  One
such effect is the coupling between the low lying magnetic and
lattice excitations to produce spin waves that interact strongly
with light by acquiring electric dipole activity from the phonons
\cite{Smol-Chupis,Katsura-DM}. These excitations, called
electromagnons \cite{Pimenov-Nature,Sushkov-Y125}, can thereby
produce contributions to the static dielectric constant which
appear in the ordered phases and that can be manipulated with an
applied magnetic field.  This may be the origin of the giant
magneto-capacitance effect observed in these materials
\cite{Hur-Dy,Kimura-113}. The magnitude of this effect is related
to the strength of the coupling between the optically active
phonons and the low lying spin waves. Therefore, further
enhancement of the effect may be achieved in materials that favor
a strong magnon-phonon coupling.

Few multiferroic systems with strong magnetoelectric effects are
known since the coexistence of proper ferroelectricity and
magnetism are usually antithetic \cite{Hill-why}. Much of the
current interest is in multiferroicity with improper
ferroelectricity which is induced by exchange striction in
magnetically ordered states with broken spatial inversion symmetry
\cite{Cheong-Mostovoy}. This produces strongly coupled
multiferroics. They are found in materials with frustrated
magnetic exchange and non-collinear spin order. One example is the
family of compounds $R$MnO$_3$ ($R$=Tb,Dy,Gd,Eu-Y), whose static
properties have been studied extensively
\cite{Kimura-113,Hemberger-Eu-Y}. In TbMnO$_3$
\cite{Kenzelmann-Tb113}, for example, the appearance of
ferroelectricity coincides with the transition of the Mn spin
system into an antiferromagnetic structure with spiral order.

The understanding of the origin of the magnetically induced
improper ferroelectricity is still evolving. The broken inversion
symmetry of the magnetic structure is only a necessary condition
for the existence of spontaneous polarization
\cite{Radaelli-symmetry,Harris-symmetry} and cannot be used to
uniquely identify the coupling mechanism. The presence of
competing exchange processes produces the ferroelectric order but
complicates the understanding. Therefore we have studied the
dynamical response which provides additional symmetry information
through the optical selection rules and gives insight about the
low lying magnetic and lattice excitations.

Pimenov et al. \cite{Pimenov-Nature,Pimenov-GdMnO3} reported the
observation of an electric dipole active magnon, electromagnon, in
the low energy electrodynamic response of Tb and GdMnO$_3$.
However this identification was not conclusive as Tb has partially
filled $f$-shells with low lying levels that can also give rise to
electric dipole excitations in this frequency range
\cite{TmFeO3-Mukhin}. A positive identification of the
electromagnon has been made in the related multiferroic compound
YMn$_2$O$_5$ \cite{Sushkov-Y125}. Since the $f$ shell in Y$^{3+}$
ion is completely empty the observed excitations arise entirely
from the Mn spin system.

However, the $R$MnO$_3$ system is attractive because it has
simpler crystalline and magnetic order which may facilitate
unravelling the complex physics of multiferroicity. Moreover, in
addition to the ambiguity due to $f$-levels, the accuracy of the
infrared (IR) data in the original reports on Tb and GdMnO$_3$ was
limited. In this letter we report IR measurements on
Eu$_{1-x}$Y$_{x}$MnO$_3$ in which Mn$^{3+}$ is the only magnetic
ion. We have chosen Eu$_{0.75}$Y$_{0.25}$MnO$_3$ because it is
close in structural, magnetic, and multiferroic properties to
TbMnO$_3$. In addition to observing electromagnons in this
compound without low energy $f$ levels, which affirms the
existence of electromagnons in the orthorhombic $R$MnO$_3$ system,
we also report a strikingly strong coupling between magnons and
phonons in Eu$_{0.75}$Y$_{0.25}$MnO$_3$ in which more than 1/2 of
the dipole oscillator strength of a phonon is transferred to a
spin excitation.

Single crystals of Eu$_{0.75}$Y$_{0.25}$MnO$_3$ were grown as
described elsewhere \cite{Rolando-Tb125}. The samples were
characterized by X-ray diffraction and dielectric measurements
in kHz range.  Our samples are ferroelectric below
$T_{FE}$=30~K with static polarization in the $a$-$c$ plane
($\bf{P_a}>\bf{P_c}$), and magnetically ordered with transition
temperature $T_N$=47~K, the magnetic structure is still
unknown. Optical measurements of reflectance and transmission
were made as a function of temperature as described elsewhere
\cite{Sushkov-Y125}. The transmission of a $a$-$b$ plane
crystal was measured at thicknesses of 1.93, 0.45, 0.080 and
0.020 mm. A second crystal was measured in $a$-$c$ plane
geometry at 1.28 mm thickness.

To extract the temperature dependence of the optical conductivity
we fit the transmission spectra with a Lorentzian model of the
dielectric constant $\varepsilon(\omega)$ for electric dipole
and/or magnetic permeability $\mu(\omega)$ for magnetic dipole
transitions as described
elsewhere\cite{Rolando-Tb125,Sushkov-Y125}:
\begin{eqnarray}
    \varepsilon(\omega) = \varepsilon_\infty+\sum_j
\frac{S_j}{\omega_{j}^2-\omega^2-\imath\omega\gamma_j}\label{eps-lorentz}
 \\
    \mu(\omega) = 1 + \sum_k
\frac{M_k}{\omega_{k}^2-\omega^2-\imath\omega\gamma_k}\label{mu-lorentz}
\end{eqnarray}
where $\varepsilon_\infty $ is the high frequency dielectric
constant, $j,k$ enumerates the oscillators, $S_j$ and $M_k$ are
spectral weights, $\omega_{j,k}$ is the resonance frequency, and
$\gamma_{j,k}$ is the damping rate.

\begin{figure}
\includegraphics[width=.9\columnwidth]{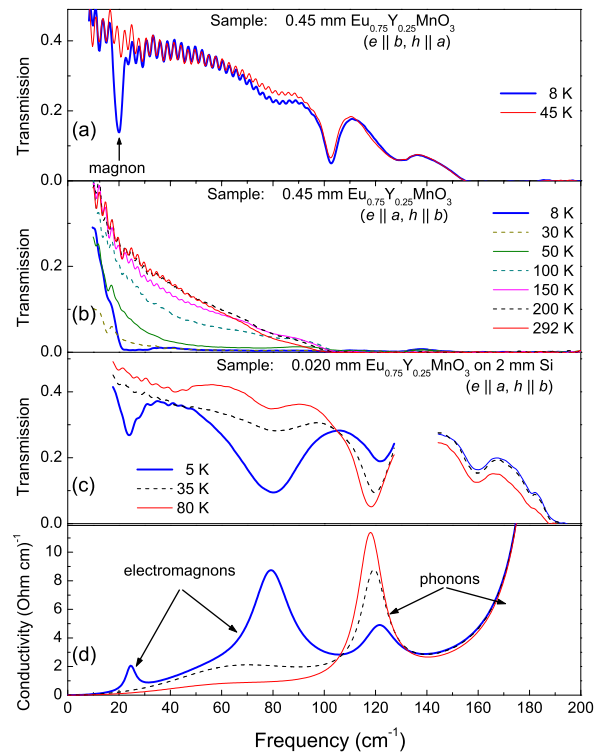}
\caption{(Color online) (a),(b), and (c) --- transmission spectra
of Eu$_{0.75}$Y$_{0.25}$MnO$_3$ in different polarization
configurations; (d)--- optical conductivity from fits of spectra
in panel (c). $e$ and $h$ are electric and magnetic fields of
light.} \label{spectra}
\end{figure}

The striking difference between figures \ref{spectra}a and
\ref{spectra}b corresponds to the magnetic dipole and electric
dipole absorptions at low temperature for the same thickness. The
transmission spectra shows that a strong low frequency absorption
in Eu$_{0.75}$Y$_{0.25}$MnO$_3$ occurs only in $e||a$ polarization
--- as was reported for TbMnO$_3$ and GdMnO$_3$
\cite{Pimenov-Nature}. In $e||c$, the IR active phonons are the
only electric dipole features observed (not shown). In the
$(e||b, h||a)$ polarization we found only one weak absorption
mode below $T_N$ (fig.~\ref{spectra}a). Fitting this mode as a
$h||a$ magnetic dipole active antiferromagnetic resonance
(AFMR) gives the values $\omega=20$~cm$^{-1}$,
$\gamma=2.3$~cm$^{-1}$, and $M=2.5$~cm$^{-2}$ at 8~K which are
typical for the AFMR \cite{Sushkov-Y125}. In this fit we used
$\varepsilon=17.5$ which was obtained by fitting $e||b$ phonon
reflectivity spectra. This resonance was also observed in the
$e||c$, $h||a$ configuration on a $a$-$c$ plane sample (not
shown).

We used a much thinner sample to quantify the $e||a$ spectra as
shown in figure \ref{spectra}c. Below $T_{FE}$, two relatively
narrow features appear; a low energy peak at 25~cm$^{-1}$ and a
broader absorption at 80~cm$^{-1}$. The gap in the
fig.~\ref{spectra}c data near 140~cm$^{-1}$ is due to absorption
in the cold quartz window. An isosbestic point (frequency of
constant absorption) is found at 105~cm$^{-1}$ which signifies
spectral weight conservation between the low frequency absorption
and the phonons. We have observed similar features at 25 and 65
cm$^{-1}$ in a polycrystal of TbMnO$_3$.

\begin{figure}
\includegraphics[width=.9\columnwidth]{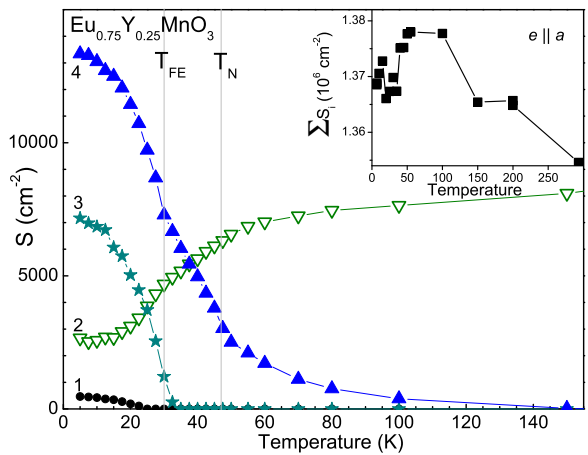}
\caption{(Color online) Spectral weight of the features below
140~cm$^{-1}$. Curves show data for: 1) 25 cm$^{-1}$ peak, 2) 120
cm$^{-1}$ phonon, 3) peak at $\sim$80~cm$^{-1}$ and, 4) total
spectral weight below 140~cm$^{-1}$ (excluding the phonon).  Inset: Total spectral weight
of the eight phonons above 140~cm$^{-1}$. } \label{sw}
\end{figure}

A broad background is absorption is observed (fig.\ref{spectra}b) in
$e||a$ (but not in the $e||b$ or $e||c$ polarizations) and persists
to temperatures as high as 150 K. In fig.\ref{sw} we show that this
background absorption grows in strength as $T$ decreases (filled
upright triangles) and this growth accelerates below $T = T_N$.  At
$T=10$ K the background accounts for approximately half of the total
low frequency oscillator strength below 140 cm$^{-1}$.

The optical conductivity obtained by fitting the transmission
spectra of fig.~\ref{spectra}c is shown in figure~\ref{spectra}d.
We used three Lorentzians to fit the spectra. Their parameters are
($\omega, \gamma, S$) at 5~K (25, 4, 467), (65, 70, 7708), (79,
17, 7506). The spectral weight of the lowest frequency peak is
comparable to the corresponding values for the electromagnons in
YMn$_2$O$_5$ and TbMn$_2$O$_5$ \cite{Sushkov-Y125}. The phonon
parameters are (122, 15, 2662) at 5~K and (118, 12, 7456) at 80~K.

The frequencies of the 25 and 80 cm$^{-1}$ peaks show very little
temperature dependence and the damping rate decreases, both below
$T_{FE}$ (not shown). The temperature dependence of the spectral
weight of the low frequency modes is shown in Figure~\ref{sw}. We
note that the phonon spectral weight begins changing significantly
around $T_N$ and shows an inflection point at $T_{FE}$, signaling
coupling to the magnetic system. It is seen that the total spectral
weight below 140~cm$^{-1}$ is not conserved; there is a net gain of
about 6,000 cm$^{-2}$. To clarify this point, we plot the total
spectral weight of the high frequency phonons in the inset. The high
frequency phonons are seen to suffer a net loss of 5,000 to
10,000~cm$^{-2}$ below $T_{N}$ which compensates for the gain of
spectral weight below 140~cm$^{-1}$ within experimental error. The
change in the phonon strength in 50--295~K range is the usual
behavior from thermal contraction. Thus, we believe that the new low
frequency modes in Eu$_{0.75}$Y$_{0.25}$MnO$_3$ are coupled to, and
acquire their optical activity from, all the phonon modes.

\begin{figure}
\includegraphics[width=.9\columnwidth]{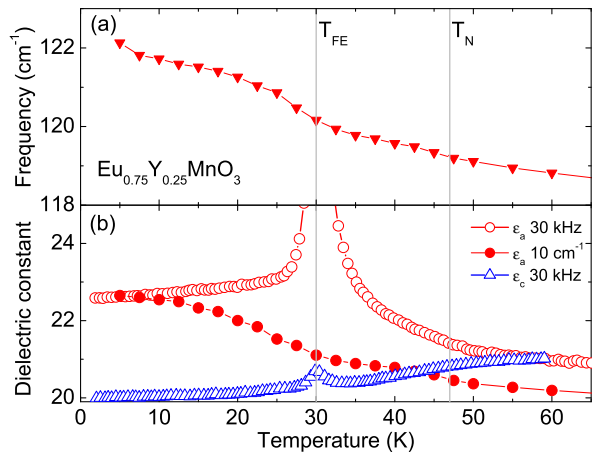}
\caption{(Color online) (a) Temperature dependence of the phonon
frequency that shows hardening at $T_{FE}$. (b) Dielectric
constant of Eu$_{0.75}$Y$_{0.25}$MnO$_3$ from fits of infrared
spectra in comparison with 30~kHz measurements.} \label{eps}
\end{figure}

Further evidence of coupling between the phonons and magnons is
visible in the temperature dependence of the phonon frequency in
Figure \ref{eps}(a). In the temperature range where phonon
hardening usually saturates we see an onset of additional
hardening at $T_{FE}$ and a smaller effect at $T_N$.
Figure~\ref{eps}(b) shows the dielectric constant of
Eu$_{0.75}$Y$_{0.25}$MnO$_3$. The peak at 30~K in the low
frequency curves ($e||a$ and $e||c$) signals the onset of the
static FE moment and is related to the dynamical response of
ferroelectric domains. The IR dielectric constant at
$\sim$10~cm$^{-1}$ only reproduces a step up in $\varepsilon_a$
which is the signature of electric dipole activity of the new
modes in $e||a$ polarization. Signatures of these new modes were
not found for $e||c$ in either transmission spectra or
$\varepsilon_c(T)$.

We now discuss the possible origin of these new IR active modes.
We clearly observe the main signature of electromagnons
($\bf{em}$)--- spectral weight transfer from phonon to
magnons\cite{Sushkov-Y125,Pimenov-Nature}. However, before
developing this interpretation further we consider several other
possibilities.

\textit{New phonons}: The activation of new phonons is expected
due to the structural distortions associated with the phase
transitions. These can arise from the lowered crystallographic
symmetry or from zone folding from a reduced Brillouin zone.  In
either case the strength of the new modes is proportional to the
square of the order parameter associated with the lattice
distortions --- $\bf{P}$. As the lattice distortions are very
small in these improper ferroelectrics, the corresponding new
phonons are also very weak. One example is provided by
TbMn$_2$O$_5$ in which two phonons that are Raman active in the PE
phase become IR active in the FE phases \cite{Rolando-Tb125}.
Their IR strength is $\sim$ 50 cm$^{-2}$ compared with 10$^4$
cm$^{-2}$ for typical IR phonons in oxides and much weaker than
the features we observe in Eu$_{0.75}$Y$_{0.25}$MnO$_3$.

\textit {Ligand field levels}: Low frequency $f$-level transitions
have been observed and reported in many rare earth compounds
\cite{TmFeO3-Mukhin}. However, our observations of new modes in
Eu$_{0.75}$Y$_{0.25}$MnO$_3$ and YMn$_2$O$_5$, cannot be explained
in this way.  Y has no $f$ levels near the Fermi level and
Eu$^{3+}$ has a ground state with J = 0, and its first excited
state is located around 300 cm$^{-1}$ above the ground state
\cite{Eu-CFL-Tovar}.

\textit{Double-well lattice model}. \citet{Golovenchits-Eu125}
interpreted a similar absorption observed in EuMn$_2$O$_5$ in
terms of two level states induced by magnetic inhomogeneities at
domain walls. This mechanism would predict unobserved variations
in strength with samples and materials. In addition, the observed
crystal structure related polarization selection rules are
incompatible with effects associated with the variable orientation
of domain boundaries.

\textit {Electromagnons}: The coupling of the magnons to the
lattice leads to mode mixing and therefore spectral weight
transfer between the electric dipole active phonons and the
magnetic dipole active magnons\cite{Katsura-DM}.  The electric
dipole activated magnon can be thought of as the Goldstone bosons
of multiferroicity \cite{Katsura-DM}.  In general the lowest order
coupling can be written as a trilinear term in the Hamiltonian
\cite{Harris-symmetry}, $H \sim uSS$, where $u$ is the lattice
displacement and $S$ is the spin variable. The form of the
interaction that can couple the $q=0$ phonon to one magnon, is $H
\sim u_0 S_{-Q}\langle S_{+Q} \rangle$, where $\langle S_{+Q}
\rangle$ corresponds to the static magnetic structure.

A quantitative comparison of experiments with theory is limited by
the lack of a theoretical treatment for realistic structures
including both symmetric and antisymmetric exchange.
\citet{Katsura-DM} have reported a theory of the $\bf{em}$ for the
case of a spin chain with cycloidal order coupled to the lattice
by Dzyaloshinskii-Moriya (DM) antisymmetric superexchange. They
predict that the $\bf{em}$ should be observed as a
$e\perp\mathbf{P}\perp q$ absorption. However the observation is
that $\textbf{em}|| a$ for Eu$_{0.75}$Y$_{0.25}$MnO$_3$ as well as
for TbMnO$_3$ and GdMnO$_3$, where the static polarization
direction and value are different for the three compounds.
Therefore the \textbf{em} in the $R$MnO$_3$ multiferroics has a
unique selection rule ($e||a$) which is not borne out by the
Katsura model. The model also predicts that the frequency of the
$\mathbf{em}$ should be lower than the AFMR, but, as shown above
in figure \ref{spectra}, we have observed the opposite. However,
the model does predict that the \textbf{em} and the AFMR are
separate modes as our result also implies.

The 80 cm$^{-1}$ feature is even more problematic within the
Katsura picture. In this case the magnon at $q=Q$ is an internal
mode in the unit cell and is nearly degenerate with the lowest
frequency phonon. However, since mode mixing has generic features,
it is interesting to examine the predictions of the Katsura model
for this nearly degenerate case. The model can produce the large
oscillator strength transfer for a large DM coupling constant but
this is accompanied by a large shift, $\sim$ 20 cm$^{-1}$, in the
optical phonon. The phonon is observed to shift to higher
frequencies, consistent with mode repulsion, but only $\sim$ 2
cm$^{-1}$.

Symmetric exchange coupling should produce similar mode mixing
behavior and in principle the coupling strength can be larger than
is expected for DM exchange. Mostovoy \cite{Mostovoy-private} has
shown that symmetric exchange coupling would produce a response of
the form $\bf{em}||\bf{P}$ for simple models.  While this is the
observed selection rule for YMn$_2$O$_5$ and TbMn$_2$O$_5$ where
$\textbf{em}||b||\bf{P}$, it is not correct for the $R$MnO$_3$
systems. Therefore, the experiments imply that the $\bf{em}$
selection rule is associated with the crystal structure and the
magnetic structure through the symmetry allowed phonon-magnon
coupling. Clearly extending the simple $\bf{em}$ models to include
more accurate depictions of the materials is an important
priority.

We note that inelastic neutron scattering can provide important
additional information. Data reported by \citet{INS-TbMnO3-Senff}
shows good agreement with the low frequency IR data of TbMnO$_3$
\cite{Pimenov-Nature}. Also S.H. Lee, et al\cite{Lee-private} have
recently reported good agreement between the sharp $\bf{em}$
features in YMn$_2$O$_5$ and the magnons observed in neutron
scattering at $q=Q$ favoring the mode mixing scenario for that
material. Mode mixing also implies the onset of magnetic dipole
activity of the phonons at $T_{FE}$.

The trilinear coupling also allows a two magnon decay of the
phonons, by the terms {$H \sim u_0 S_{q}S_{-q}$}.  The
corresponding frequency dependent phonon self energy can produce
modifications of the phonon absorption line shape related to the
two magnon ($q_1=-q_2$) density of states. As the two magnon
density of states is broad with no narrow spectral features this
process is unlikely to produce the observed relatively narrow
electromagnon absorption. However, the two magnon process is a
good candidate for understanding the broad background absorption
observed below as well as above $T_N$ in the $e||a$ polarization.
The gradual decrease of this background signal with temperature
(Fig. \ref{sw}) above $T_N$ suggests that it originates from
magnetic fluctuations. Since there are no long lived magnons for
$T>T_N$ this background arises from the coupling of the phonon to
the dynamic fluctuations of the magnetic system in the
paramagnetic phase. Confirming evidence for this interpretation
would be the observation of short range magnetic order in these
materials above $T_N$ by inelastic neutron scattering.

Summarizing the above discussion we interpret our observations as
follows: 1) the well defined absorption peaks at 25 and 80
cm$^{-1}$ arise from mode mixed phonon-magnon excitations ---
electromagnons, and 2) there exists colossal coupling between
magnon and phonons that leads to the spectacular loss of spectral
weight of the 120 cm$^{-1}$ phonon. We assert that electromagnon
excitations provide a powerful new window into the physics of
multiferroics and that a treatment of the magnon-phonon
interaction that includes the crystal and magnetic structure may
allow a better understanding of the exchange coupling mechanism
responsible for the exotic magneto-electric properties of these
materials.

We thank C. Broholm, A.B. Harris, D.~Khomskii and M.~Mostovoy for
useful discussions. This work was supported in part by the
National Science Foundation MRSEC under Grant No. DMR-0520471.

\bibliography{Eu-Y113-electromagnons}
\end{document}